\documentclass{kapproc} 

\usepackage{procps} 

\usepackage[dvips]{graphicx}

\upperandlowercase

\kluwerbib 

\begin{document}

\articletitle{Searches for high redshift galaxies \\
using gravitational lensing}

\author{J. Richard$^{1}$, R. Pello$^{1}$,  J.-P. Kneib$^{1,2}$, D. Schaerer$^{1,3}$, M.R. Santos$^{2}$, R. Ellis$^{2}$}

\affil{$^{1}$O.M.P., Laboratoire d'Astrophysique, UMR 5572, 14 Avenue E. Belin, 31400 Toulouse, France\\
$^{2}$California Institute of Technology, 105-24 Caltech, Pasadena, CA 91125, USA\\
$^{3}$Geneva Observatory, 51 Ch. des Maillettes, CH-1290 Sauverny, Switzerland\\
}
\email{jrichard@ast.obs-mip.fr}


\begin{abstract}
We present different methods used to identify high redshift ($z>5$) objects
in the high-magnification regions of lensing galaxy clusters, taking
advantage of very well constrained lensing models. The research procedures
are explained and discussed. The detection of emission lines in the
optical/NIR spectra, such as Lyman-alpha, allows us to determine the redshift
of these sources. Thanks to the lensing magnification, it is possible to
identify and to study more distant or intrinsically fainter objects with
respect to standard field surveys.
\end{abstract}

\begin{keywords}
galaxies: formation, evolution, high-redshift, luminosity function,\\
clusters : lensing ---cosmology: observations
\end{keywords}

\section{Introduction}
The main purpose of looking at high redshift ($z>5$) objects is
to get constraints about the nature and the formation epoch of the
first sources in the Universe. The advent of 8-10 m class telescopes, such as VLT and Keck, has opened up this field of study. Moreover, the use of clusters of
 galaxies as gravitational telescopes can help a lot for this.  
Strong lensing effect in clusters has already enabled the detection of one of the most distant galaxies known up to now (Hu et al. 2002), thanks to the gravitational magnification and despite the decrease in effective area of the survey.\\ \indent
We present here two methods aimed at the detection of lyman-alpha sources
behind galaxy clusters : first a spectroscopic search along the critical
lines of clusters, and then a photometric selection technique for very
low metallicity starbursts (the so-called Population III objects, Loeb \& 
Barkana 2001), using ultra-deep near infrared imaging.

\section {Critical lines survey}
\begin{figure}[ht]
\begin{minipage}{7.5cm}
\includegraphics[width=7cm]{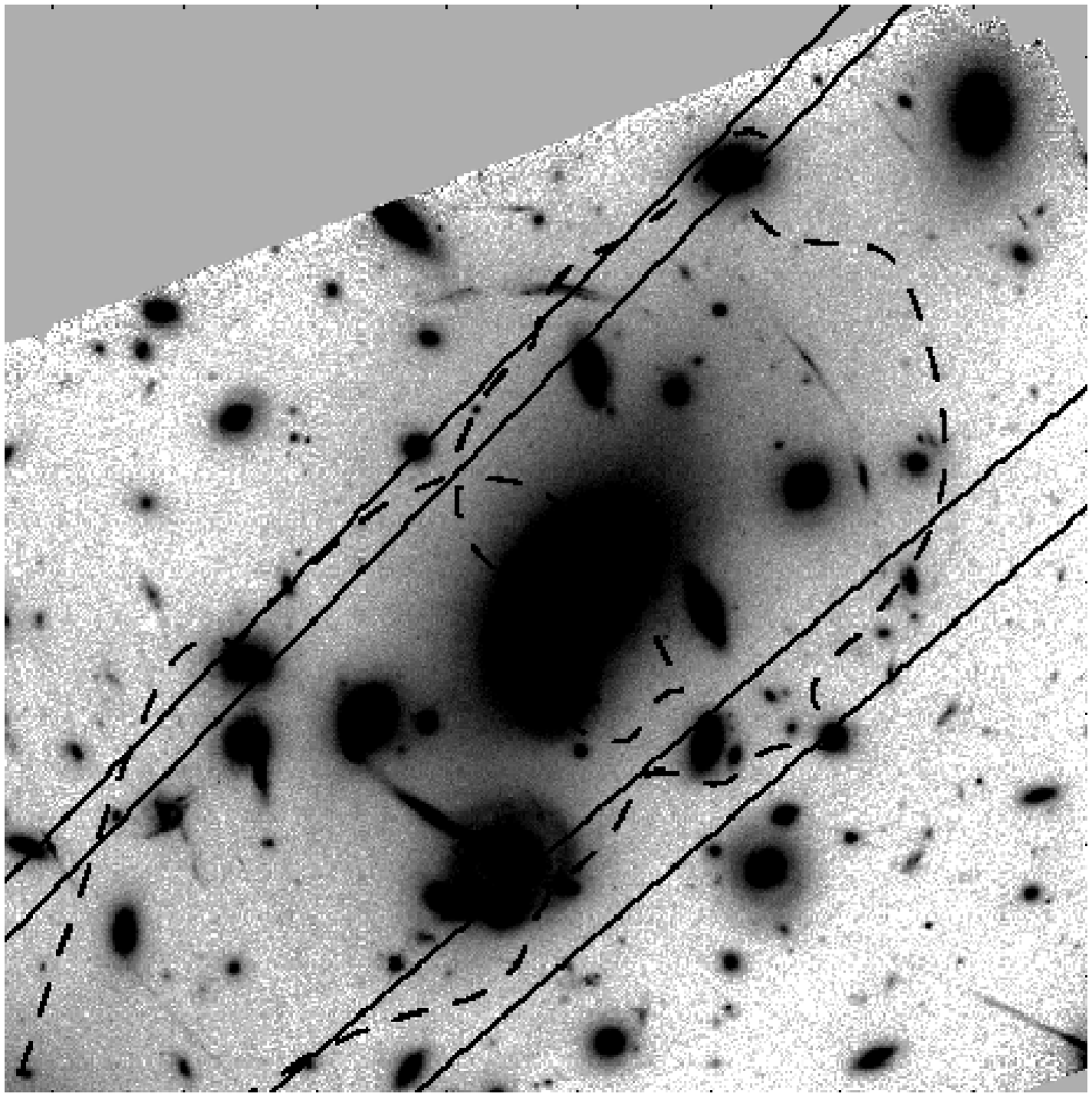}
\end{minipage}
\begin{minipage}{4cm}
\includegraphics[angle=180,width=3.8cm]{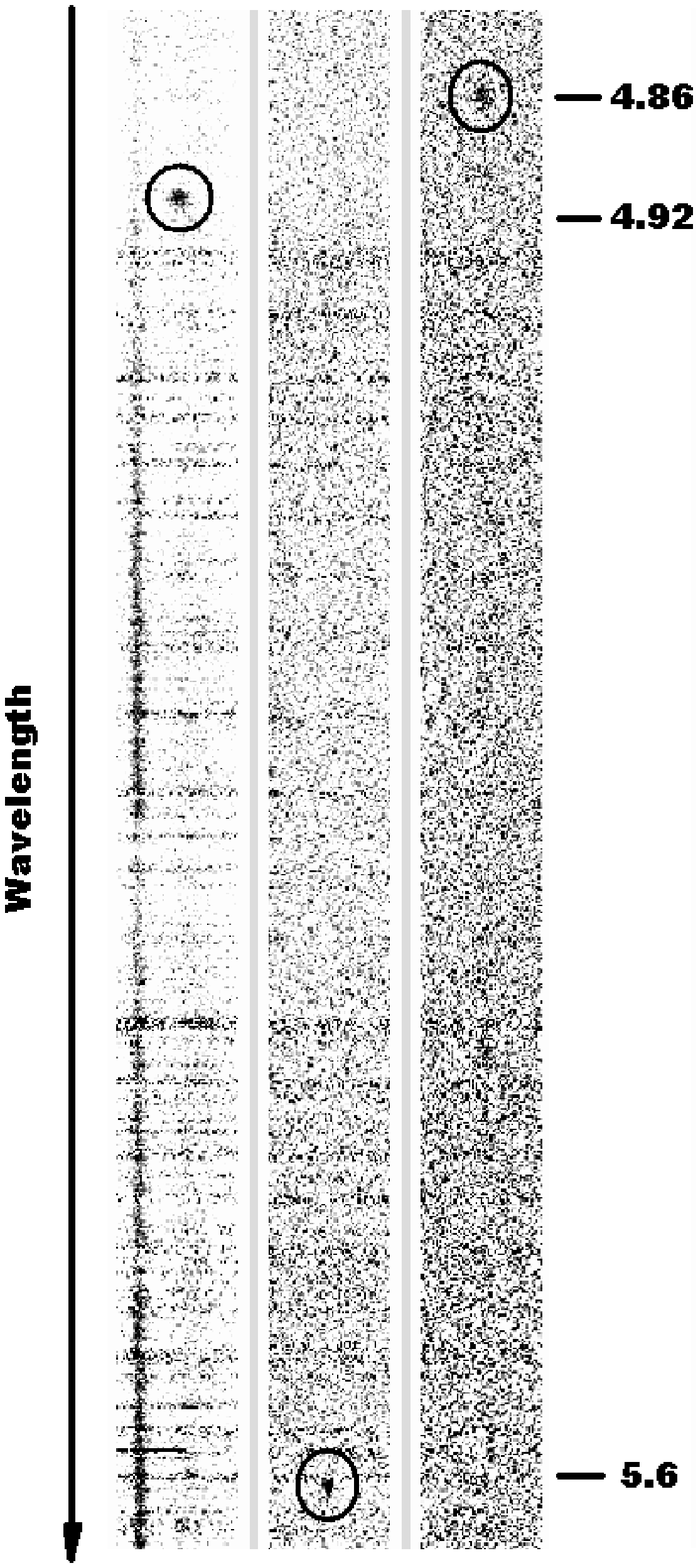}
\end{minipage}
\caption{\label{slit}On the left: zoom on the center part of the galaxy cluster Abell 2218. The critical lines for z=5 are shown as dashed lines, and the two regions scanned by the survey as rectangles. On the right: composite spectra of the three $Ly\alpha$ emitters (circled) found at $z\sim5-6$.}
\end{figure}

Using the LRIS spectrograph at Keck, we searched for $Ly\alpha$ emitters at redshift 2.5 to 6.8 in the most magnified parts of a sample of lensing clusters, selected for having well-constrained mass models. We scanned the regions located near the critical lines (Figure \ref{slit}), defined as the lines of infinite magnification for a given redshift, using a 175"-long slit. Half of the area covered (4.2 arcmin$^2$) is at least magnified by a factor of 10 at $z=5$.  \indent\par 
We systematically looked for every single emission line in the spectra, and we confirmed $Ly\alpha$ candidates using HST images available for these clusters, optical photometry, and further spectroscopy at higher resolution, using ESI at Keck that can easily resolve the [OII] doublet, thus preventing this contamination. We identified 12 $Ly\alpha$ candidates, three of them lying in the redshift range $\sim 4.6-5.6$. One is a double image at $z\sim5.6$ which was analysed with more details by Ellis et. al (2001). The two-dimensionnal LRIS spectra, showing $Ly\alpha$ emission lines, are presented in figure \ref{slit}.
\indent\par
Thanks to the strong lensing magnification, these results can give us constraints on the luminosity function of emitters at $4.6 < z < 5.6$ with $Ly\alpha$ luminosity $10^{40}<L<10^{42}\,erg/s$, which is a depth that was not reached by other surveys of $Ly\alpha$ emitters or Lyman Break Galaxies. This will be presented in Santos et al. (2003, \textit{ApJ} submitted).
\section {Looking for Population III objects}
Recent models by D. Schaerer (2002, 2003) for the Spectral Energy Distributions (SED) of Population III objects show that they may be currently observable using 8 - 10 m telescopes, at the limits of conventional spectroscopy. The identification of such objects should be possible thanks to their very strong  emission lines, mainly $Ly\alpha$ and HeII $\lambda$1640. In order to find these objects, the colors predicted by the same models can allow us to define a color-color region in the near-infrared diagram (J-H) vs (H-K'), where we can pick up candidates (Fig. \ref{cand}). By doing simulations with existing models, we found that we should not be contaminated by stars or $z<8$ galaxies, even in the case of important redenning.\indent\par
As a first test of these selection criteria, we did very deep imaging (limiting magnitudes of J=25, H=24.5, K'=24, Vega system), with ISAAC on VLT, of two lensing clusters, taking advantage of the lensing magnification to help us detecting these faint objects.
\section{Preliminary results}
\begin{figure}[ht]
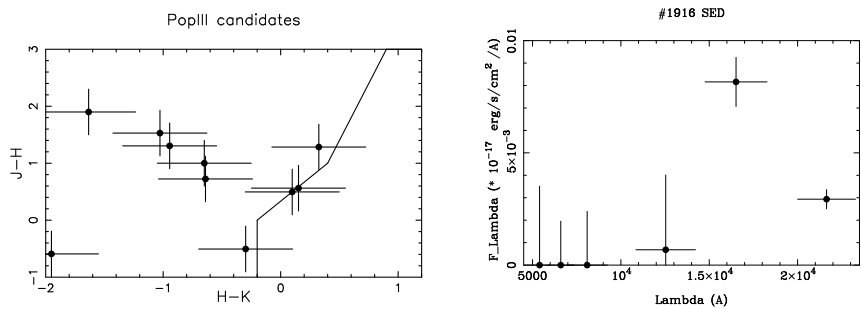

\begin{minipage}{6cm}
\centerline{\mbox{\includegraphics[angle=270,width=5.5cm]{cc.ps}}}
\end{minipage}
\begin{minipage}{6cm}
\centerline{\mbox{\includegraphics[angle=270,width=5cm]{sed.ps}}}
\end{minipage}
\caption{\label{cand}Left: location, on the NIR color-color diagram (J-H) vs (H-K'), of the candidates found with the typical photometric errors (Vega system). The selection region for Population III objects is delimited by a solid line. Right : example of SED, combining infrared and optical photometry, for one of the candidates. Photometric redshift gives $z\sim 9$.}
\end{figure}

We selected several ($\sim$ 10) candidates per cluster, satisfying our selection criteria in the near-IR, and being undetected on available optical images. These objects have the expected magnitudes and SEDs of $8<z<10$ Population III objects (Figure \ref{cand}), and are magnified by 2 to 4 magnitudes thanks to the strong lensing effect. We used a modified version of the photometric software \textbf{hyperz} (Bolzonella et al. 2000) to find the redshift distribution probability of our candidates with the spectra models quoted above.\indent\par
\begin{figure}[ht]
\centerline{\mbox{\includegraphics[angle=270,width=9.55cm]{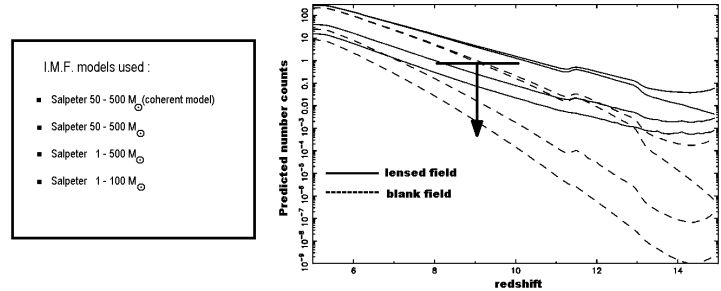}}}
\caption{\label{nc}Number counts of PopIII objects per interval of 0.1 in z, expected to have K'<24 in the ISAAC fov. The different curves correspond to different IMF models. The values obtained with gravitational lensing (solid curves) are boosted by a factor of 10 at $z \sim 8 -10$ regarding the one expected for a blank field (dashed curves). Overplotted is the upper limit of our survey.}
\end{figure}

As a preliminary result, we can try to compare the number of Population III objects per redshift that was expected to be detected in our field with the upper limit corresponding to our candidates, using a simple model of dark-matter halos distribution (Press \& Schechter, 1974), and 4 different models of IMF for PopIII galaxies (Figure \ref{nc}). Furthermore, we can estimate the efficiency of using strong lensing in this field by plotting the expected number counts in a blank field of same size and depth. We find that lensing is more efficient at high redshifts ($z>8$), and that the number of candidates we found is consistent with some of the models we used.
\section {Conclusions} 
The use of gravitational lensing is efficient to detect more distant or intrinsically fainter galaxies lying behind galaxy clusters : we can have constraints on luminosity functions at fainter scales, and the expected number of primordial objects in a cluster field is boosted at high redhsifts.\indent\par
Even if the candidates we found with our selection criteria are very faint, we should try to perform spectroscopy of the best ones with present day facilities. The detection of these sources is one of the major science cases for the next generation NIR instruments, like EMIR/GTC, KMOS/VLT or KIRMOS/Keck.

\end{document}